\PassOptionsToPackage{unicode}{hyperref}
\PassOptionsToPackage{hyphens}{url}
\PassOptionsToPackage{dvipsnames,svgnames,x11names}{xcolor}
\documentclass[
]{article}
\usepackage{amssymb}
\usepackage{lmodern}
\usepackage{iftex}
\usepackage{setspace}
\usepackage{amsmath}
\usepackage{titlesec}
\usepackage{macros}
\usepackage{natbib}
\ifPDFTeX
  \usepackage[T1]{fontenc}
  \usepackage[utf8]{inputenc}
  \usepackage{textcomp} 
\else 
  \usepackage{unicode-math}
  \defaultfontfeatures{Scale=MatchLowercase}
  \defaultfontfeatures[\rmfamily]{Ligatures=TeX,Scale=1}
\fi
\IfFileExists{upquote.sty}{\usepackage{upquote}}{}
\IfFileExists{microtype.sty}{
  \usepackage[]{microtype}
  \UseMicrotypeSet[protrusion]{basicmath} 
}{}
\makeatletter
\@ifundefined{KOMAClassName}{
  \IfFileExists{parskip.sty}{%
    \usepackage{parskip}
  }{
    \setlength{\parindent}{0pt}
    \setlength{\parskip}{6pt plus 2pt minus 1pt}}
}{
  \KOMAoptions{parskip=half}}
\makeatother
\usepackage{xcolor}
\usepackage{longtable,booktabs,array}
\usepackage{calc} 
\usepackage{etoolbox}
\makeatletter
\patchcmd\longtable{\par}{\if@noskipsec\mbox{}\fi\par}{}{}
\makeatother
\IfFileExists{footnotehyper.sty}{\usepackage{footnotehyper}}{\usepackage{footnote}}
\makesavenoteenv{longtable}
\usepackage{graphicx}
\makeatletter
\def\maxwidth{\ifdim\Gin@nat@width>\linewidth\linewidth\else\Gin@nat@width\fi}
\def\maxheight{\ifdim\Gin@nat@height>\textheight\textheight\else\Gin@nat@height\fi}
\makeatother
\setkeys{Gin}{width=\maxwidth,height=\maxheight,keepaspectratio}
\makeatletter
\def\fps@figure{htbp}
\makeatother
\newlength{\cslhangindent}
\newlength{\csllabelwidth}
\newlength{\cslentryspacingunit} 
\newenvironment{CSLReferences}[2] 
 {
  \setlength{\parindent}{0pt}
  \ifodd #1
  \let\oldpar\par
  \def\par{\hangindent=\cslhangindent\oldpar}
  \fi
  \setlength{\parskip}{#2\cslentryspacingunit}
 }%
 {}
\usepackage{calc}

\ifLuaTeX
\usepackage[bidi=basic]{babel}
\else
\usepackage[bidi=default]{babel}
\fi
\babelprovide[main,import]{american}

\def\languageshorthands#1{}
\ifLuaTeX
  \usepackage{selnolig}  
\fi
\IfFileExists{bookmark.sty}{\usepackage{bookmark}}{\usepackage{hyperref}}
\IfFileExists{xurl.sty}{\usepackage{xurl}}{} 
\renewcommand{\vec}[1]{\mathbf{#1}}

\hypersetup{
  pdftitle={PyPLUTO: a data analysis Python package for the PLUTO code},
  pdfauthor={Giancarlo Mattia, Daniele Crocco, 
             David Melon Fuksman, Matteo Bugli, 
             Vittoria Berta, Eleonora Puzzoni, 
             Andrea Mignone, Bhargav Vaidya},
  pdflang={en-US},
  colorlinks=true,
  linkcolor={Maroon},
  filecolor={Maroon},
  citecolor={Blue},
  urlcolor={Blue},
  pdfcreator={LaTeX via pandoc}}

\title{\textbf{PyPLUTO: a data analysis Python package for the PLUTO code}}

\usepackage[affil-it]{authblk}
\usepackage{orcidlink}
\author[1,2%
  \ensuremath\mathparagraph]{Giancarlo Mattia%
    \,\orcidlink{0000-0003-1454-6226}\,%
    }
\author[3%
  ]{Daniele Crocco%
    \,\orcidlink{0009-0001-4284-9446}\,%
    }
\author[1%
  ]{David Melon Fuksman%
    \,\orcidlink{0000-0002-1697-6433}\,%
    }
\author[3,4,5,6%
  ]{Matteo Bugli%
    \,\orcidlink{0000-0002-7834-0422}\,%
    }
\author[3,6%
  ]{Vittoria Berta%
    \,\orcidlink{0000-0001-6305-6931}\,%
    }
\author[7%
  ]{Eleonora Puzzoni%
    \,\orcidlink{0009-0009-5314-348X}\,%
    }
\author[3,6%
  ]{Andrea Mignone%
    \,\orcidlink{0000-0002-8352-6635}\,%
    }
\author[8%
  ]{Bhargav Vaidya%
    \,\orcidlink{0000-0001-5424-0059}\,%
    }

\affil[1]{Max-Planck-Institut f\"{u}r Astronomie, K\"{o}nigstuhl 17, Heidelberg, 69117, Germany}
\affil[2]{INFN, Sezione di Firenze, Via G. Sansone 1, Sesto Fiorentino (FI), 50019, Italy}
\affil[3]{Dipartimento di Fisica, Università di Torino, Via P. Giuria 1, Torino, 10125, Italy}
\affil[4]{Institut d’Astrophysique de Paris, UMR 7095, CNRS \& Sorbonne Universit\'e, 98 bis boulevard Arago, 75014 Paris, France}
\affil[5]{Universit\'e Paris-Saclay, Universit\'e Paris Cit\'e, CEA, CNRS, AIM, Gif-sur-Yvette, 91191, France}
\affil[6]{INFN, Sezione di Torino, Via P. Giuria 1, Torino, 10125, Italy}
\affil[7]{Observatoire de la C\^ote d’Azur, Laboratoire Lagrange, Bd de l’Observatoire, CS 34229, 06304 Nice cedex 4, France}
\affil[8]{Department of Astronomy, Astrophysics and Space Engineering, Indian Institute of Technology, Khandwa Road, Simrol, Indore, 453552, India}

\affil[$\mathparagraph$]{Corresponding author (mattia@mpia.de)}

\date{\today}

\usepackage[margin=1.05in]{geometry}
\usepackage[scaled]{beramono}
 
\definecolor{darkgreen}{rgb}{0.05, 0.3, 0.1}
\let\oldtexttt\texttt
\renewcommand{\texttt}[1]{\oldtexttt{\textcolor{darkgreen}{#1}}}
\usepackage{caption}

\begin{document}
\maketitle


\hypertarget{summary}{%
\section{Summary}\label{summary}}
In recent years, numerical simulations have become indispensable for addressing complex astrophysical problems.
The so-called magnetohydrodynamics (MHD) framework represents a key tool for investigating the dynamical evolution of astrophysical plasmas. This formalism consists of a set of partial differential equations (\protect\hyperlink{ref-Chiuderi2015}{Chiuderi \& Velli, 2015}) that enforce the conservation of mass, momentum, and energy, along with Maxwell’s equations for the evolution of the electromagnetic fields.
Due to the high nonlinearity of the MHD equations (regardless of their specifications, e.g., classical/relativistic or ideal/resistive), a general analytical solution is not possible, making numerical approaches crucial. 
Numerical simulations usually produce large sets of data files, and their scientific analysis relies on dedicated software tools designed for data visualization (\protect\hyperlink{ref-Ahrens2005}{Ahrens et al., 2005}; \protect\hyperlink{ref-Childs2012}{Childs et al., 2012}).
However, to encompass all code output features, specialized tools focusing on the numerical code may represent a more versatile and integrated solution.

\begin{figure}
  \centering
  \includegraphics[width=0.97\textwidth]{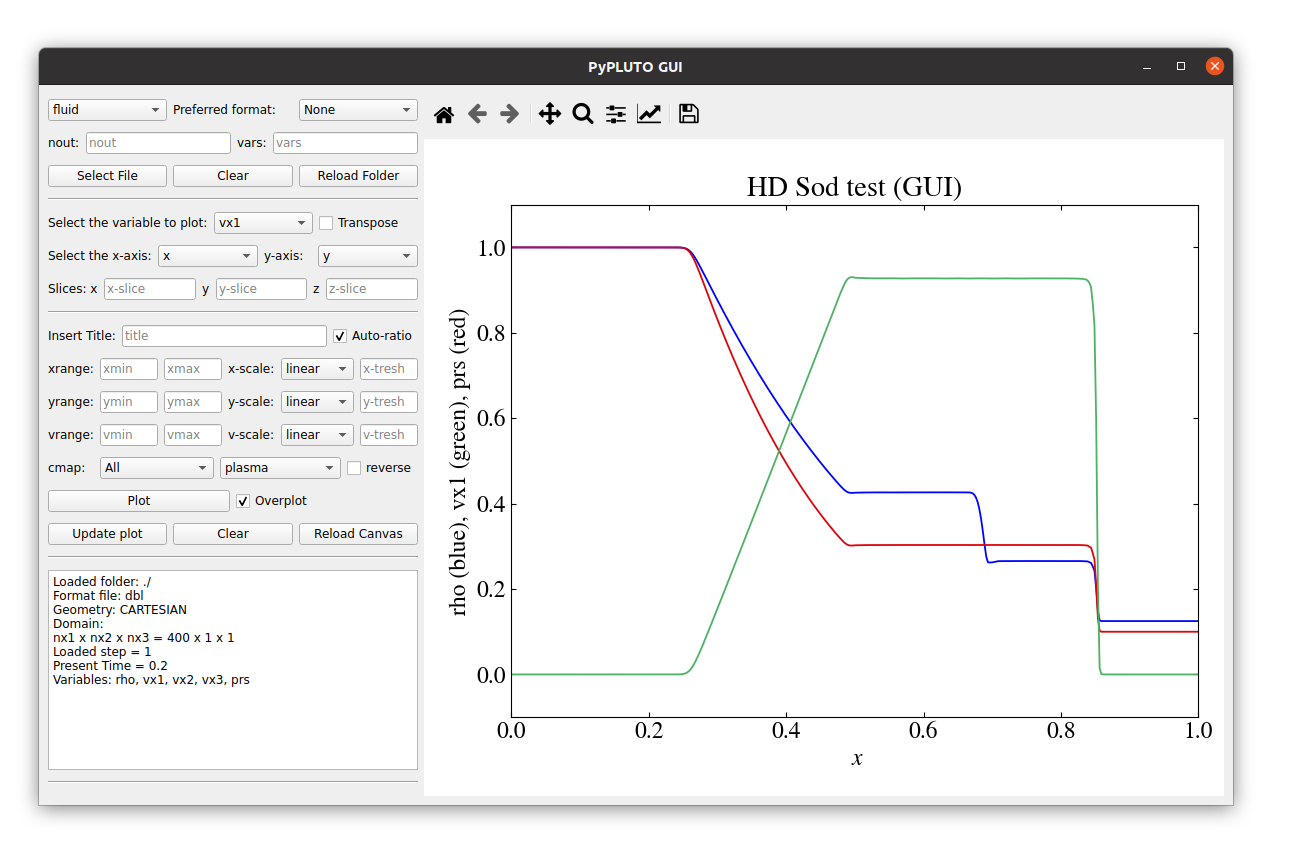}
  \caption{\footnotesize Interactive visualization of shock tube test results (i.e., density, pressure, and velocity profiles) with the GUI.
  }
  \label{fig::pyplutogui}
\end{figure}

Here, we present PyPLUTO, a Python package tailored for efficient loading, manipulation, and visualization of outputs produced with the PLUTO\footnote{
\url{https://plutocode.ph.unito.it}} code 
(\protect\hyperlink{ref-Mignone2007}{Mignone et al., 2007}; 
 \protect\hyperlink{ref-Mignone2012a}{Mignone, Zanni et al., 2012}).
 PyPLUTO uses memory mapping to optimize data loading and provides general data manipulation and visualization routines. PyPLUTO also supports the particle modules of the PLUTO code, enabling users to load and visualize particles, such as cosmic rays (\protect\hyperlink{ref-Mignone2018}{Mignone et al., 2018}), Lagrangian 
(\protect\hyperlink{ref-Vaidya2018}{Vaidya et al., 2018}), or dust particles
(\protect\hyperlink{ref-Mignone2019}{Mignone et al., 2019}), from hybrid simulations.
A dedicated graphical user interface (GUI, shown in Fig. \ref{fig::pyplutogui}) simplifies the generation of single-subplot figures, making PyPLUTO a powerful yet user-friendly toolkit for astrophysical data analysis.

\hypertarget{statement-of-need}{%
\section{Statement of need}\label{statement-of-need}}

The PLUTO code (\protect\hyperlink{ref-Mignone2007}{Mignone et al., 2007}) is a widely-used, freely-distributed computational fluid dynamics code designed to solve the classical and (special) relativistic MHD equations in different geometries and spatial dimensions. The original code is written in C (while the upcoming GPU version provides a complete C++ rewrite\footnote{\url{https://plutocode.ph.unito.it/pluto-gpu.html}}) and it contains several numerical methods adaptable to different contexts. 
Data post-processing is a crucial step in analyzing the results of any numerical simulation. Other packages addressing related needs (e.g  plutoplot\footnote{\url{https://github.com/Simske/plutoplot}}), provide valuable functionality for working with PLUTO data, including loading and visualization.
However, they may not support all data formats or offer integration for data manipulation and advanced plotting tasks. In this work, we present PyPLUTO, a Python package designed to efficiently load, manipulate, and visualize the output from the PLUTO code.

The initial version of PyPLUTO (written by Bhargav Vaidya) is available here\footnote{\url{https://github.com/coolastro/pyPLUTO}}. 
PyPLUTO has since been completely rewritten and is now maintained at a new repository\footnote{\url{https://github.com/GiMattia/PyPLUTO}}.
The package retains its core strengths while offering user-friendly methods for generating publication-quality plots with high customization. In addition to its enhanced flexibility, PyPLUTO offers strong computational efficiency, enabling the rapid handling of large datasets typical of state-of-the-art numerical simulations. Through this balance between customization, performance, and ease of use, PyPLUTO represents a key tool to effectively communicate scientific results while minimizing the effort required for post-processing.

\hypertarget{main-features}{%
\section{Main features}\label{main-features}}

PyPLUTO is a package written for Python 3.10 or later with the additions of external packages
such as NumPy (\protect\hyperlink{ref-Harris2020}{Harris et al., 2020}), Matplotlib 
(\protect\hyperlink{ref-Hunter2007}{Hunter et al., 2007}), SciPy 
(\protect\hyperlink{ref-Virtanen2020}{Virtanen et al., 2020}), pandas 
(\protect\hyperlink{ref-McKinney2010}{McKinney, 2010}), h5py 
(\protect\hyperlink{ref-Collette2013}{Collette, 2013}), and PyQT6\footnote{\url{https://www.riverbankcomputing.com/software/pyqt/intro}}.
The package, which can be installed through pip, primarily consists of three main classes:

\begin{itemize}
    \item The \texttt{Load} class loads and manipulates the PLUTO output files containing fluid-related quantities.

    \item The \texttt{LoadPart} class loads and manipulates the PLUTO output files containing particle-related quantities.

    \item The \texttt{Image} class produces and handles the graphical windows and the plotting procedures.

\end{itemize}

Additionally, a separate \texttt{PyPLUTOApp} class launches a GUI able to load and plot 1D and 2D data in a single set of axes. 
PyPLUTO has been implemented to be supported by Windows, macOS, and Linux, through both standard scripts and more interactive tools (e.g., IPython or Jupyter). 
The style guidelines follow the PEP8\footnote{\url{https://peps.python.org/pep-0008/}} conventions for Python codes, enforced through the Black package \protect\hyperlink{ref-Langa2020}{Langa et al., 2020}, and focus on clarity and code readability. 
 Finally, by leveraging the capabilities of the Sphinx package\footnote{\url{http://sphinx-doc.org/sphinx.pdf}}, PyPLUTO features extensive docstrings, providing a helpful reference for both users and developers.

\section{Benchmark examples}

PyPLUTO provides a set of benchmarks that are immediately accessible after the package is installed. These consist of test problems that can be applied to relevant astrophysical applications and showcase the full range of PyPLUTO’s features. Here we report two examples demonstrating the package’s capabilities.

\hypertarget{sec::ex1}{%
\subsection{Disk-Planet Interaction}\label{sec::ex1}}

\begin{figure}
  \centering
  \includegraphics[width=0.97\textwidth]{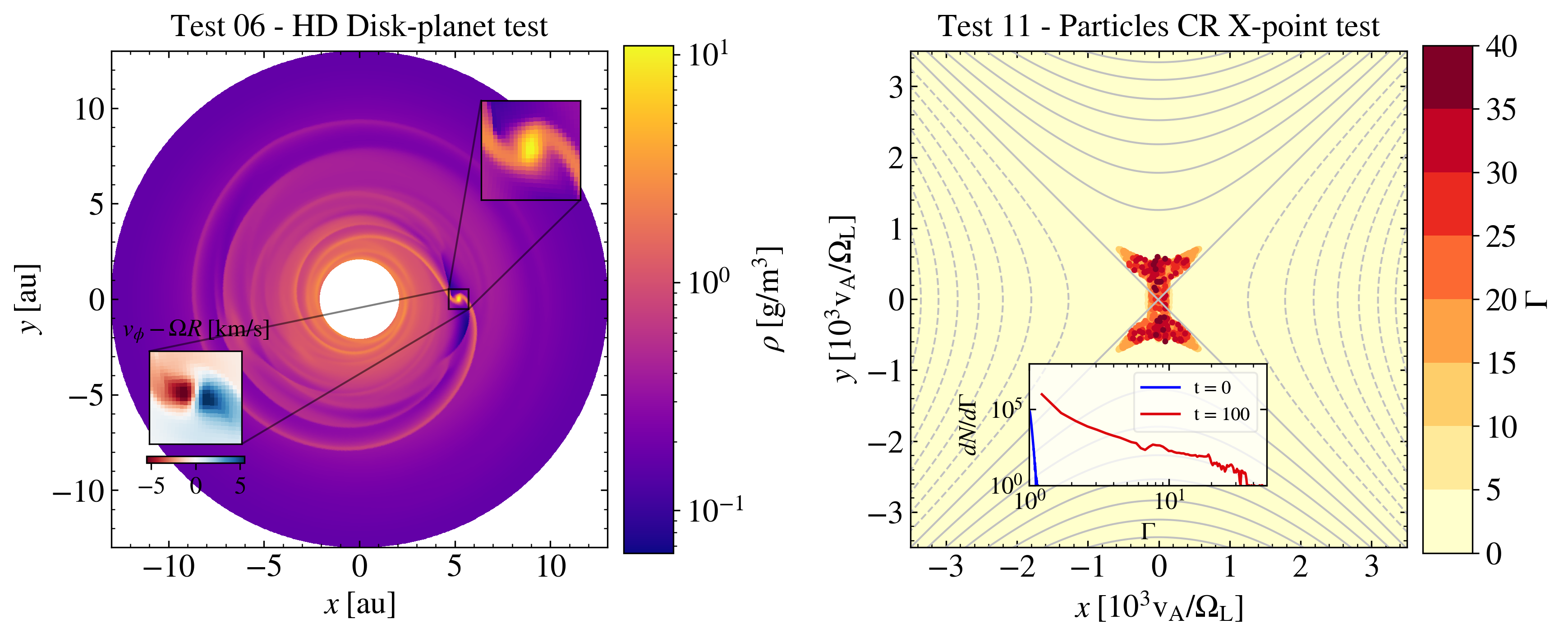}
    \caption{\footnotesize Left panel: example of inset zooms of the planet region of the disk-planet test problem. The main plot and the right zoom show the density on a logarithmic scale, while the left zoom highlights the toroidal velocity on a linear scale. 
    Right panel: example of an X-point region with magnetic field lines overlaid (as contour lines of the vector potential, solid lines). The main plot shows the test-particle distribution, color-coded by velocity magnitudes, while the inset plot displays the particle energy spectrum at the beginning (in blue) and end (in red) of the simulation.}
  \label{fig::plots}
\end{figure}

This test simulates the interaction of a planet embedded in a disk 
(\protect\hyperlink{ref-Mignone2012b}{Mignone, Flock et al., 2012}) and represents an ideal scenario for understanding the formation and evolution of planetary systems.
In particular, the formation of spiral density waves and disk gaps represents some key observational signatures of planet formation and planet-disk interaction 
(\protect\hyperlink{ref-MelonFuksman2021}{Melon Fuksman et al., 2021},
\protect\hyperlink{ref-Muley2024}{Muley et al., 2024}).
In the left panel of Fig. \ref{fig::plots}, we show an adaptation of Figure 10 of \protect\hyperlink{ref-Mignone2012b}{Mignone, Flock et al., 2012}, featuring two separate zoom-ins around the planet's location.

\begin{itemize}
\item The first zoom (upper-right subplot) shows an enlarged view of the density distribution using the same color map and logarithmic scale as the global plot.

\item The second zoom (lower-left subplot) highlights the changes in toroidal velocity due to the planet’s presence by employing a different color map (to enhance the sign change) and a linear color scale.
\end{itemize}

These zoomed views offer deeper insights into the physical processes at play and demonstrate the utility of PyPLUTO for analyzing complex astrophysical systems.

\hypertarget{sec::ex2}{%
\subsection{Particles Accelerated near an X-point}\label{sec::ex2}}

This test problem examines particle acceleration near an X-type magnetic reconnection region (\protect\hyperlink{ref-Puzzoni2021}{Puzzoni et al., 2021}). 
In the last decades, magnetic reconnection (\protect\hyperlink{ref-Mattia2023}{Mattia et al., 2023}, \protect\hyperlink{ref-Bugli2024}{Bugli et al., 2024}) has proven to be a key physical process to explain the population of non-thermal particles in solar flares, relativistic outflows, and neutron star magnetospheres. 
This sort of test provides valuable insights into particle acceleration mechanisms in high-energy astrophysical environments by enabling the investigation of particle trajectories and energy distribution near the X-point.

In the right panel of Fig. \ref{fig::plots} we show an adaptation of the top panel of Figure 13-14 from \protect\hyperlink{ref-Mignone2018}{Mignone et al., 2018}. 
The main plot displays the distribution of test particles, color-coded by their velocity magnitudes, with magnetic field lines overlaid as solid and dashed lines. 
The inset panel shows the energy spectrum at the initial ($t = 0$, in blue) and final ($t = 100$, in red) time.
In this scenario, the absence of a guide field ($\vec{E} \cdot \vec{B} = 0$) results in a symmetric distribution along the y-axis from the combined effects of the gradient, curvature, and $\vec{E} \times \vec{B}$ drifts in the vicinity of the X-point, where the electric field is the strongest.

This plot provides a clear visual representation of particle motion and energy changes, demonstrating how PyPLUTO can be used to investigate complex processes such as particle acceleration in astrophysical sources.

\hypertarget{ongoing-research}{%
\section{Ongoing research using PyPLUTO}\label{ongoing-research}}

Research applicable with PyPLUTO includes the development of numerical algorithms (\protect\hyperlink{ref-Mattia2022a}{Mattia \& Mignone, 2022}, 
\protect\hyperlink{ref-Berta2024}{Berta et al., 2024}, 
\protect\hyperlink{ref-MelonFuksman2025}{Melon Fuksman et al. 2025}) and numerical simulations of astrophysical objects, such as jets 
(\protect\hyperlink{ref-Mattia2022b}{Mattia \& Fendt, 2022}, 
\protect\hyperlink{ref-Mattia2023}{Mattia et al. 2023}, 
\protect\hyperlink{ref-Mattia2024}{Mattia et al. 2024},
\protect\hyperlink{ref-Costa2025}{Costa et al. 2025},
\protect\hyperlink{ref-Sciaccaluga2025}{Sciaccaluga et al. 2025}), star clusters
(\protect\hyperlink{ref-Haerer2025}{Härer et al. 2025})
and protoplanetary disks 
(\protect\hyperlink{ref-MelonFuksman2024a}{Melon Fuksman et al. 2024a}, 
\protect\hyperlink{ref-MelonFuksman2024b}{Melon Fuksman et al. 2024b}), as well as physical processes, such as particle acceleration (\protect\hyperlink{ref-Wang2024}{Wang et al., 2024}) and magnetic reconnection (\protect\hyperlink{ref-Bugli2025}{Bugli et al., 2025}).

\hypertarget{conclusions-and-future-perspectives}{%
\section{Conclusion and Future Perspectives}\label{conclusions-and-future-perspectives}}

The PyPLUTO package is designed as a powerful yet flexible tool to facilitate the data analysis and visualization of the output from PLUTO simulations, focusing on user friendliness while allowing the necessary customization to produce publication-quality figures. To overcome current limitations and further enhance the package’s capabilities, particular focus will be devoted to:

\begin{itemize}
    \item introducing specific routines for rendering 3D data to provide users with tools for visualizing volumetric data;

    \item supporting interactive visualization and comparison of multiple simulation outputs, allowing the users to track temporal evolution directly with the GUI; and

    \item expanding the graphical interface to support particle data, including dynamic visualization of particle distributions and trajectories.

\end{itemize}
Alongside these improvements, PyPLUTO development will focus on encompassing the latest features of the PLUTO code, such as new adaptive mesh refinement strategies and extensions to more general metric tensors. 
PyPLUTO is a public package that can be downloaded alongside the CPU and GPU versions of the PLUTO code\footnote{\url{https://gitlab.com/PLUTO-code/gPLUTO}}.
Regular updates will be released with improvements and bug fixes.
Additionally, a repository\footnote{\url{https://github.com/GiMattia/PyPLUTO}} containing the PyPLUTO development versions will be available for users who wish to exploit the code's latest features in advance.

\hypertarget{acknowledgments}{%
\section{Acknowledgments}\label{acknowledgments}}
The authors thank the reviewers for improving this work with valuable comments and suggestions.
G. Mattia thanks L. Del Zanna and M. Flock for the discussions on data visualization and the Data Science Department of the Max Planck Institute for Astronomy for helping with Python and Matplotlib.

The authors thank Simeon Doetsch for their insights on memory-mapping techniques and Deniss Stepanovs and Antoine Strugarek for contributing to previous PyPLUTO versions throughout the years. The authors thank Agnese Costa, Alberto Sciaccaluga, Alessio Suriano, Asmita Bhandare, Dhruv Muley, Dipanjan Mukherjee, Jieshuang Wang, Jacksen Narvaez, Lucia Haerer, Prakruti Sudarshan, Stefano Truzzi, and Stella Boula for testing the module while it was under development. M. Bugli acknowledges the support of the French Agence Nationale de la Recherche (ANR), under grant ANR-24-ERCS-0006 (project BlackJET). This project has received funding from the European Union’s Horizon Europe research and innovation programme under the Marie Skłodowska-Curie grant agreement No 101064953 (GR-PLUTO), and from the European High Performance Computing Joint Undertaking (JU) and Belgium, Czech Republic, France, Germany, Greece, Italy, Norway, and Spain under grant agreement No 101093441 (SPACE).

\hypertarget{references}{%
\section*{References}\label{references}}
\addcontentsline{toc}{section}{References}

\hypertarget{refs}{}
\begin{CSLReferences}{0.7}{0.7}

\leavevmode\vadjust pre{\hypertarget{ref-Ahrens2005}{}}%
Ahrens, J., Geveci, B. \& Law, C. (2005),
{ParaView}: An End-User Tool for Large Data Visualization,
\emph{Visualization Handbook, Butterworth-Heinemann}, \emph{ISBN}~978-0123875822

\leavevmode\vadjust pre{\hypertarget{ref-Berta2024}{}}%
Berta, V., Mignone, A., Bugli, M., \& Mattia, G. (2024),
A 4$^{\text{th}}$-order accurate finite volume method for ideal classical and special relativistic MHD based on pointwise reconstructions,
\emph{\jcp}, \emph{499}, 112701.
\url{https://doi.org/10.1016/j.jcp.2023.112701}

\leavevmode\vadjust pre{\hypertarget{ref-Bugli2025}{}}%
Bugli, M., Lopresti, E. F., Figueiredo, E., Mignone, A., Cerutti, B., Mattia, G., Del Zanna, L., Bodo, G., \& Berta, V. (2025),  
Relativistic reconnection with effective resistivity: I. Dynamics and reconnection rate,  
\emph{\aap}, \emph{693}, A233.  
\url{https://doi.org/10.1051/0004-6361/202452277}

\leavevmode\vadjust pre{\hypertarget{ref-Childs2012}{}}%
Childs, H., Brugger, E., Whitlock, B., Meredith, J., Ahern, S., Pugmire, D., Biagas, et al. (2012),
VisIt: An End-User Tool For Visualizing and Analyzing Very Large Data,
\emph{High Performance Visualization--Enabling Extreme-Scale Scientific Insight},
\url{https://doi.org/10.1201/b12985}

\leavevmode\vadjust pre{\hypertarget{ref-Chiuderi2015}{}}%
Chiuderi, C., Velli, M. (2015),
Basics of Plasma Astrophysics,
\url{https://doi.org/10.1007/978-88-470-5280-2}

\leavevmode\vadjust pre{\hypertarget{ref-Collette2013}{}}%
Collette, A. (2013),
Python and HDF5,
\emph{O'Reilly}

\leavevmode\vadjust pre{\hypertarget{ref-Costa2025}{}}%
Costa, A., Bodo, G., Tavecchio, F., Rossi, P., Coppi, P., Sciaccaluga, A., \& Boula, S. (2025),  
How do recollimation-induced instabilities shape the propagation of hydrodynamic relativistic jets?,  
\emph{arXiv e-prints}, arXiv:2503.18602.  
\url{https://doi.org/10.48550/arXiv.2503.18602}

\leavevmode\vadjust pre{\hypertarget{ref-Haerer2025}{}}%
H{\"a}rer, L., Vieu, T., \& Reville, B. (2025),  
Stellar-wind feedback and magnetic fields around young compact star clusters: 3D magnetohydrodynamics simulations,  
\emph{\aap}, \emph{698}, A6.  
\url{https://doi.org/10.1051/0004-6361/202554057}

\leavevmode\vadjust pre{\hypertarget{ref-Harris2020}{}}%
Harris, C. R., Millman, K. J., van der Walt, S. J., Gommers, R., Virtanen, P., Cournapeau,
 D., Wieser, E., et al. (2020),
Array programming with NumPy,
\emph{\nat}, \emph{585(7825)}, 357–362.
\url{https://doi.org/10.1038/s41586-020-2649-2}

\leavevmode\vadjust pre{\hypertarget{ref-Hunter2007}{}}%
Hunter, J. D. (2007),
Matplotlib: A 2D Graphics Environment,
\emph{Computing in Science and Engineering}, \emph{9(3)}, 90-95.
\url{https://doi.org/10.1109/MCSE.2007.55}

\leavevmode\vadjust pre{\hypertarget{ref-Langa2020}{}}%
Langa, Ł., \& contributors to Black. (2020),
Black: The uncompromising Python code formatter,
\emph{Black documentation}, \emph{https://black.readthedocs.io/en/stable/}.
\url{https://doi.org/10.3847/1538-4357/aabccd}

\leavevmode\vadjust pre{\hypertarget{ref-Mattia2023}{}}%
Mattia, G., Del Zanna, L., Bugli, M., Pavan, A., Ciolfi, R., Bodo, G., \& Mignone, A. (2023),
Resistive relativistic MHD simulations of astrophysical jets,
\emph{\aap}, \emph{679}, A49.
\url{https://doi.org/10.1051/0004-6361/202347126}

\leavevmode\vadjust pre{\hypertarget{ref-Mattia2024}{}}%
Mattia, G., Del Zanna, L., Pavan, A., \& Ciolfi, R. (2024),  
Magnetic dissipation in short gamma-ray-burst jets: I. Resistive relativistic MHD evolution in a model environment,  
\emph{\aap}, \emph{691}, A105.  
\url{https://doi.org/10.1051/0004-6361/202451528}

\leavevmode\vadjust pre{\hypertarget{ref-Mattia2022b}{}}%
Mattia, G., \& Fendt, C. (2022),  
Jets from Accretion Disk Dynamos: Consistent Quenching Modes for Dynamo and Resistivity,  
\emph{\apj}, \emph{935(1)}, 22.  
\url{https://doi.org/10.3847/1538-4357/ac7d59}

\leavevmode\vadjust pre{\hypertarget{ref-Mattia2022a}{}}
A comparison of approximate non-linear Riemann solvers for Relativistic MHD,
\emph{\mnras}, \emph{510(1)}, 481-499.
\url{https://doi.org/10.1093/mnras/stab3373}

\leavevmode\vadjust pre{\hypertarget{ref-McKinney2010}{}}%
McKinney, W. (2010),
Data Structures for Statistical Computing in Python,
\emph{Proceedings of the 9th Python in Science Conference}, \emph{56–61}.
\url{https://doi.org/10.25080/Majora-92bf1922-00a}

\leavevmode\vadjust pre{\hypertarget{ref-MelonFuksman2024a}{}}%
Melon Fuksman, J. D., Flock, M., \& Klahr, H. (2024),  
Vertical shear instability in two-moment radiation-hydrodynamical simulations of irradiated protoplanetary disks. I. Angular momentum transport and turbulent heating,  
\emph{\aap}, \emph{682}, A139.  
\url{https://doi.org/10.1051/0004-6361/202346554}

\leavevmode\vadjust pre{\hypertarget{ref-MelonFuksman2024b}{}}%
Melon Fuksman, J. D., Flock, M., \& Klahr, H. (2024),  
Vertical shear instability in two-moment radiation-hydrodynamical simulations of irradiated protoplanetary disks. II. Secondary instabilities and stability regions,  
\emph{\aap}, \emph{682}, A140.  
\url{https://doi.org/10.1051/0004-6361/202346555}

\leavevmode\vadjust pre{\hypertarget{ref-MelonFuksman2025}{}}%
Melon Fuksman, D., Flock, M., Klahr, H., Mattia, G., \& Muley, D. (2025),  
Multidimensional half-moment multigroup radiative transfer: Improving moment-based thermal models of circumstellar disks,  
\emph{\aap}, \emph{701}, A97.  
\url{https://doi.org/10.1051/0004-6361/202554994}

\leavevmode\vadjust pre{\hypertarget{ref-MelonFuksman2021}{}}%
Melon Fuksman, J. D., Klahr, H., Flock, M., \& Mignone, A. (2021),
A Two-moment Radiation Hydrodynamics Scheme Applicable to Simulations of Planet Formation in Circumstellar Disks,
\emph{\apj}, \emph{906(2)}, 78.
\url{https://doi.org/10.3847/1538-4357/abc879}

\leavevmode\vadjust pre{\hypertarget{ref-Mignone2007}{}}%
Mignone, A., Bodo, G., Massaglia, S., Matsakos, T., Tesileanu, O., Zanni, C., \& Ferrari, A. (2007),
PLUTO: A Numerical Code for Computational Astrophysics,
\emph{\apjs}, \emph{170(1)}, 228-242.
\url{https://doi.org/10.1086/513316}

\leavevmode\vadjust pre{\hypertarget{ref-Mignone2018}{}}%
Mignone, A., Bodo, G., Vaidya, B., \& Mattia, G. (2018),
A Particle Module for the PLUTO Code. I. An Implementation of the MHD-PIC Equations,
\emph{\apj}, \emph{859(1))}, 13.
\url{https://doi.org/10.3847/1538-4357/aabccd}

\leavevmode\vadjust pre{\hypertarget{ref-Mignone2012b}{}}%
Mignone, A., Flock, M., Stute, M., Kolb, S. M., \& Muscianisi, G. (2012),
A conservative orbital advection scheme for simulations of magnetized shear flows with the PLUTO code,
\emph{\aap}, \emph{545}, A152.
\url{https://doi.org/10.1051/0004-6361/201219557}

\leavevmode\vadjust pre{\hypertarget{ref-Mignone2019}{}}%
Mignone, A., Flock, M.,  \& Vaidya, B. (2019),
A Particle Module for the PLUTO Code. III. Dust,
\emph{\apjs}, \emph{244(2)}, 38.
\url{https://doi.org/10.3847/1538-4365/ab4356}

\leavevmode\vadjust pre{\hypertarget{ref-Mignone2012a}{}}%
Mignone, A. , Zanni, C., Tzeferacos, P., van Straalen, B., Colella, P., \& Bodo, G. (2012),
The PLUTO Code for Adaptive Mesh Computations in Astrophysical Fluid Dynamics,
\emph{\apjs}, \emph{198(1)}, 7.
\url{https://doi.org/10.1088/0067-0049/198/1/7}

\leavevmode\vadjust pre{\hypertarget{ref-Muley2024}{}}%
Muley, D., Melon Fuksman, J. D., Klahr, H. (2024),
Three-temperature radiation hydrodynamics with PLUTO: Thermal and kinematic signatures of accreting protoplanets,
\emph{\aap}, \emph{687}, A213.
\url{https://doi.org/10.1051/0004-6361/202449739}

\leavevmode\vadjust pre{\hypertarget{ref-Puzzoni2021}{}}%
Puzzoni, E., Mignone, A., \& Bodo, G. (2021),
On the impact of the numerical method on magnetic reconnection and particle acceleration - I. The MHD case,
\emph{\mnras}, \emph{ 508(2)},  2771–2783.
\url{https://doi.org/10.1093/mnras/stab2813}

\leavevmode\vadjust pre{\hypertarget{ref-Sciaccaluga2025}{}}%
Sciaccaluga, A., Costa, A., Tavecchio, F., Bodo, G., Coppi, P., \& Boula, S. (2025),  
The polarization of the synchrotron radiation from a recollimated jet: Application to high-energy BL Lacs,  
\emph{\aap}, \emph{699}, A296.  
\url{https://doi.org/10.1051/0004-6361/202554490}

\leavevmode\vadjust pre{\hypertarget{ref-Tol2021}{}}%
Tol, P. (2021),
Colour Schemes,
\emph{SRON/EPS/TN/09-002},  \emph{3.2(1)} 1-18
\url{https://personal.sron.nl/~pault/data/colourschemes.pdf}

\leavevmode\vadjust pre{\hypertarget{ref-Vaidya2018}{}}%
Vaidya, B., Mignone, A., Bodo, G., Rossi, P. \& Massaglia, S. (2018),
A Particle Module for the PLUTO Code. II. Hybrid Framework for Modeling Nonthermal Emission from Relativistic Magnetized Flows,
\emph{\apj}, \emph{865(2)},  144.
\url{https://doi.org/10.3847/1538-4357/aadd17}

\leavevmode\vadjust pre{\hypertarget{ref-Virtanen2020}{}}%
Virtanen, P., Gommers, R., Oliphant, T. E., Haberland, M., Reddy, T., Cournapeau, D., Burovski, E., et al. (2020),
SciPy 1.0: Fundamental algorithms for scientific computing in Python,
\emph{Nature Methods}, \emph{17(2)}, 261–272.
\url{https://doi.org/10.1038/s41592-019-0686-2}

\leavevmode\vadjust pre{\hypertarget{ref-Wang2024}{}}%
Wang, J.-S., Reville, B., Rieger, F. M., \& Aharonian, F. A. (2024),  
Acceleration of Ultra-high-energy Cosmic Rays in the Kiloparsec-scale Jets of Nearby Radio Galaxies,  
\emph{\apjl}, \emph{977(1)}, L20.  
\url{https://doi.org/10.3847/2041-8213/ad9589}

\end{CSLReferences}

\end{document}